\documentclass[twocolumn,amsmath,amssymb]{revtex4-2}
\usepackage{multirow}
\usepackage{array}
\usepackage{amsmath}
\usepackage{graphicx}% Include figure files
\usepackage{dcolumn}% Align table columns on decimal point
\usepackage{bm}% bold math
\usepackage{verbatim}
\usepackage{float}
\usepackage{physics}
\usepackage{xcolor}
\usepackage{color,soul}
\DeclareMathAlphabet \mathbfcal{OMS}{cmsy}{b}{n}

\begin{document}

\title{Three-level Spaser System: a Semi-Classical Analysis}

\author{Rupesh Ghimire$^1$}
\email{ghrupesh23@gmail.com}
\affiliation{%
	Center for Nano-Optics (CeNO) and Department of Physics and Astronomy, Georgia State University, Atlanta, Georgia 30303
}%
\author{Dalton .C. Hunley$^2$ }
\email{dhunley1@student.gsu.edu}
\affiliation{%
	Center for Nano-Optics (CeNO) and Department of Physics and Astronomy, Georgia State University, Atlanta, Georgia 30303
}%
\author{Fatemeh Nematollahi$^2$}
\email{nematolahi.buph@gmail.com }
\affiliation{%
	UCSanDiego, 9500 Gilman Drive, La Jolla, CA 92093-0021
}%
\author{Sayed .J. Hossaini$^2$}
\email{shossaini1@student.gsu.edu}
\affiliation{%
	Center for Nano-Optics (CeNO) and Department of Physics and Astronomy, Georgia State University, Atlanta, Georgia 30303
}%
\author{Suresh Gnawali$^2$}
\email{sgnawali1@student.gsu.edu}
\affiliation{%
	Center for Nano-Optics (CeNO) and Department of Physics and Astronomy, Georgia State University, Atlanta, Georgia 30303
}%
\author{Vadym Apalkov$^1$}
\email{vapalkov@gsu.edu}
\affiliation{%
	Center for Nano-Optics (CeNO) and Department of Physics and Astronomy, Georgia State University, Atlanta, Georgia 30303
}%
%\affiliation{%
% Department of Physics and Astronomy, Georgia State University, Atlanta, Georgia 30303
%}%

\date{\today}
\begin{abstract}
	We theoretically study a nanospaser system, which consists of a spherical silver nanoparticle embedded inside a sphere composed of dye molecules. The gain of the system, dye molecules, is described by a three-level model, where the transition frequency between the lowest two energy levels is close to the surface plasmon frequency of the nanosphere. Contrary to a two-level model of spaser, the three-level model takes into account finite relaxation time between the high energy levels of the gain medium. These 
	relaxation processes affect both the spaser threshold and the number of generated plasmons in the continuous wave regime. While for a two-level model of a spaser the number of generated plasmons has a linear dependence on the gain, for the three-level model this dependence becomes quadratic.
	
\end{abstract}
\pacs{}
\maketitle

\section{Introduction}
%The world today centers on miniaturization. 
Scaling down electronic and optical systems to harness their optimum efficiency has been a goal of today's scientific and industrial research. Among the research fields, working in this direction, nanoplasmonics plays an important role due to its unique possibilities and broad applications \cite{stockman11_Nanoplasmonics_applications,stockman11_nanoplasmonics_present_past_glimpse,dong20_Recent_application_nanoplasmonics}.  In nanoplasmonics, the light is confined within a sub-wavelength scale, which results in a strong enhancement of the optical field. Some of the notable application of nanoplasmonics are in the areas of near-field optics \cite{kawata09_near_field_imaging,kholmicheva18_nearfield_applications}, bio-sensing \cite{anker10_biosensing_plasmonic,kim2019nanoplasmonic,zhang2017nanoplasmonic},  surface plasmon-based photo-detectors \cite{photodetector1,tang08_germanium_photodetector}, spaser\cite{stockman20_brief_history,azzam20_ten_years_spaser,premaratne17_Theory_spaser}
and many others. Spaser (surface plasmon amplification by stimulated emission of radiation), which has experienced a vast development over the last decade, plays a special role in this list. It was introduced in the early 2000's \cite{bergman03_SP_amplification_Stimulated} by  David. J. Bergman and Mark. I . Stockman  \cite{aizpurua21_Evangelist,boltasseva21_stockman_nature_knight},  and throughout the time has paved its way up as a miniature source of spectrally tunable stimulated emission \cite{li05_SP_amplification_nanolenses,stockman10_Nanoscale_quantum_generator}. Apart from the mainstream research, spaser has found numerous applications in different areas such as opto-electronic systems\cite{mark18_CMOS,khajavikhan12_Thresholdless_nanoscale,ning19_semiconductor_nanolasers,shane14_metallo_dielectric_nanolasers,shen18_hyperbolic_metacavity,ding15_bandwidth_modulation,dolores-calzadilla17_Waveguide_coupled_nanopillar}, sensing in biological and chemical agents\cite{cheng18_defect_cavity_for_sensing,wang17_Nanolaser_aqueous_solution,wang12_Imaging_organs} and also as a biological probe\cite{galanzha17_Biological_probe,fan14_Optofluidic_biolaser} in disease therapeutics and diagnostics.

The whole idea of spaser is based on the existence of localized surface plasmons, which are characterized by a high concentration of optical energy within a nanoscale 
range\cite{bergman92_macroscopically_inhomogeneous_media}. 
Such strong concentration of optical field combined with stimulated emission \cite{bergman03_SP_amplification_Stimulated} process allow to design a nanoscale laser - spaser. Different variations of such nanoscopic lasers were proposed theoretically and realized experimentally. The first of the type\cite{zheludev08_Lasing_spaser} was introduced in 2008 and was based on an array of plasmonic resonators. A year later, a spaser, based on Localized Surface Plasmons Resonance\cite{willets07_LSPR,mayer11_LSPR}(LSPR), was demonstrated experimentally\cite{noginov09_Spaser_based_nanolaser}, which contained a gold sphere embedded inside a dye. The same design has been also been studied for a cancer diagnosis and treatment by Ganzala et al\cite{galanzha17_Biological_probe} in 2017

The design of spaser based on a gain nanorod placed near plasmonic metal with the dimensions in micrometers has been considered in Ref. \cite{oulton09_Plasmon_lasers,motavas19_low_threshold_nanolaser}. Recently,
the topological spasers of type I and II were introduced\cite{wu20_Topological_spaser,ghimire20_Topological_nanospaser,ghimire21_TMDC_based}, 
where the spaser dynamics is protected by nontrivial topological properties of either a plasmonic system\cite{wu20_Topological_spaser} or a gain medium\cite{ghimire20_Topological_nanospaser,ghimire21_TMDC_based}.

% tried to study the topological aspect of a Spaser by introducing chirality in the metal and gain separately. Topological Spaser\cite{wu20_Topological_spaser} and Topological nanospaser\cite{ghimire20_Topological_nanospaser,ghimire21_TMDC_based}, both discuss an additional degree of freedom ( angular momentum), which is responsible for topological charge in the plasmons. This enables one to control the generation of plasmons with a particular rotation in application of a circularly-polarized light. Altogether, Spaser has been a promising area of research with unique prospects and designs. 
\begin{figure}
	\begin{center}
		\includegraphics[width=0.45\textwidth]{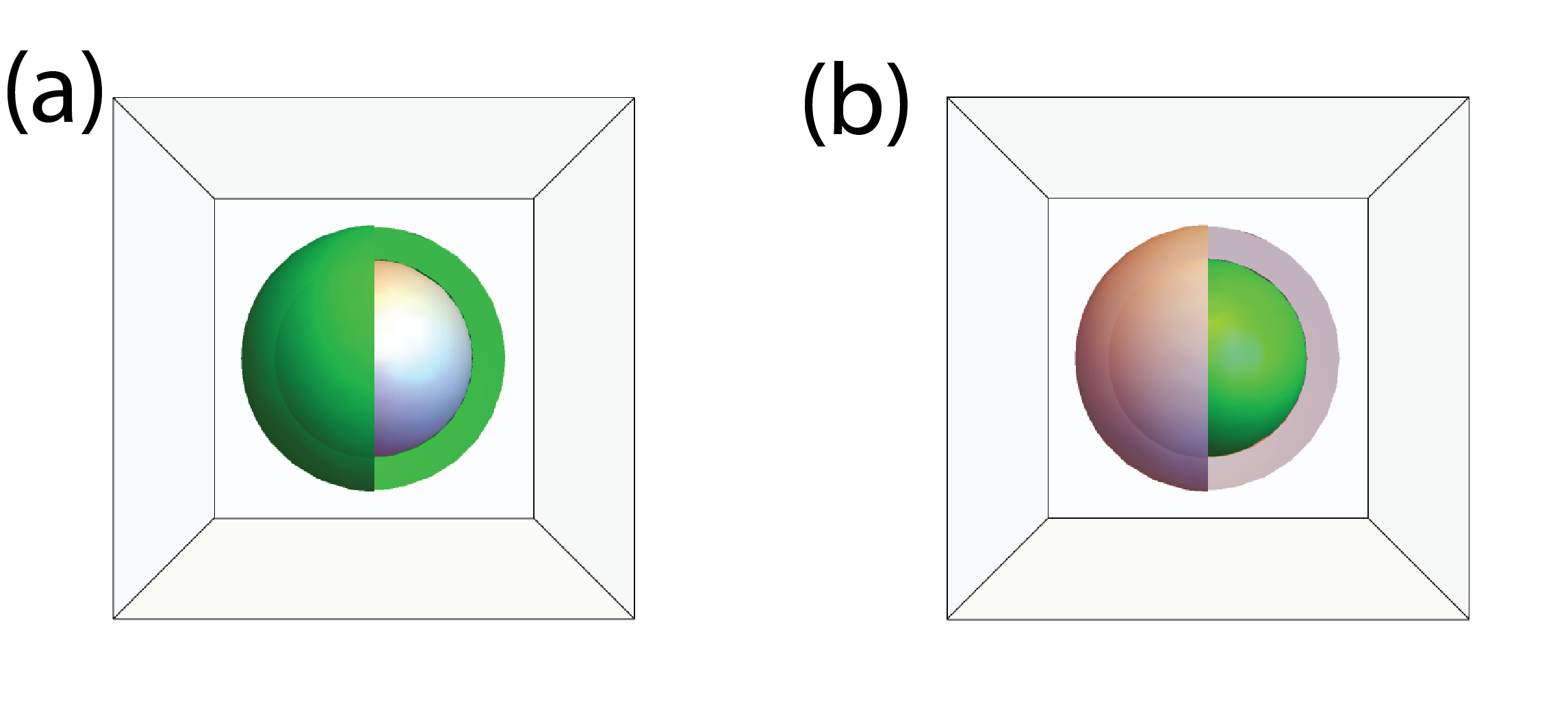}
	\end{center}
	\caption{ Schematic illustration of two geometries of spaser: (a) a metal nanosphere surrounded by a gain medium (shown by green) (b) a gain medium placed inside a metal nanoshell. }
	\label{two_setup}
\end{figure}

Spaser can be understood as a nanoplasmonic counterpart of a normal laser\cite{maiman1960stimulated}. It consists of two major components: a metal resonator and a gain medium. Usually, the metal is silver due to its low non-radiative losses, however, gold or aluminum can also be used for the spaser operating at high frequency. The gain medium is usually dye molecules,  with the gap close to the SP frequency. Apart from dyes, materials\cite{liu13_three_band_model,with_2019_morphology,nematollahi20_Topological_resonance_Weyl} with non-linear effects that exhibit topological resonance, have also been studied as a suitable gain.  Additionally, the system may also be placed in an appropriate dielectric environment, which can be used to adjust the SP frequency of a nanosphere. %A metal resonator supports the plasmonic modes and gain supplies energy to these modes undergoing population inversion. Spaser overcomes the limitation of Laser as the localization of SP modes is in nanoscale compared to electric field modes in a Laser which ranges in micro-scale.

The two basic possibilities of how a gain medium can be introduced into a spaser system are shown in Fig. \ref{two_setup}: i) gain encapsulating the solid nanosphere ii) gain placed inside a metal nanoshell. Below, we consider only the first case when the gain is placed outside the solid silver metal nanosphere as shown in Fig. \ref{two_setup}a. Additionally, this system is placed in water which will further help to adjust the LSP frequency($\omega_{\mathrm{sp}}$).

%Here we consider dipole mode $l=1$ for our calculations considering quadruple effects are small enough to neglect. 
Theoretical modeling\cite{stockman10_Nanoscale_quantum_generator} of such nanolaser\cite{noginov09_Spaser_based_nanolaser} is mostly in agreement with the experimental results, apart from the behavior of the spaser at a large value of electric pumping. 
In the present paper, we are using the same theoretical model\cite{stockman10_Nanoscale_quantum_generator} of spaser with some modifications, which can address the unexplained behavior. Namely, we consider a three-level model for the gain while in Ref.\ \cite{stockman10_Nanoscale_quantum_generator} only two levels were introduced for the gain medium.

\begin{figure}
	\begin{center}
		\includegraphics[width=0.45\textwidth]{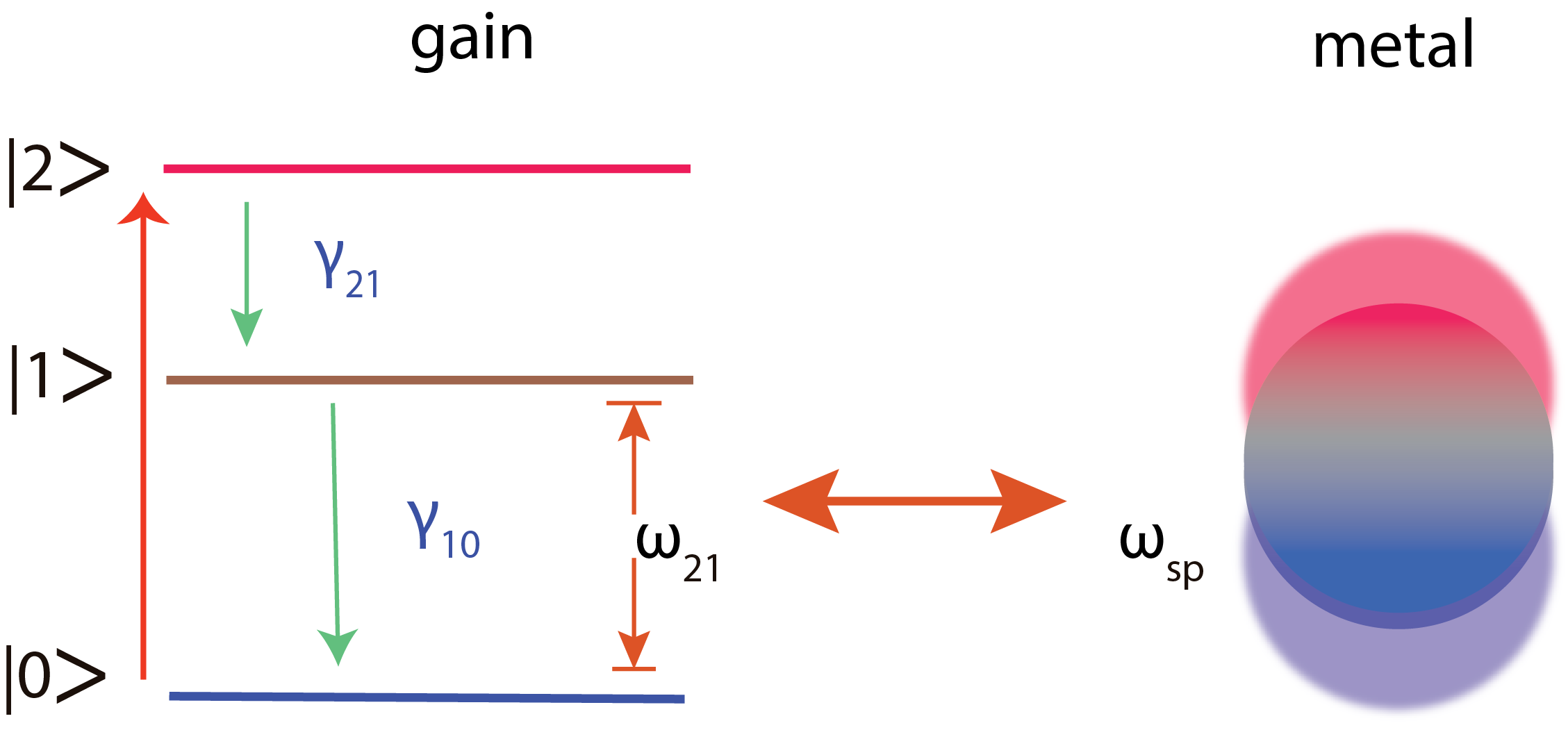}
	\end{center}
	\caption{Schematics of energy levels of dye (gain medium) and a silver sphere. Here,  $|0>$, $|1>$, and $|2>$ are the three levels of the dye (gain) with the corresponding populations $n_1, n_2$ and $n_2$. External laser pulse pumps the system and excites the gain medium from the ground state $|0>$ to the second excited one $|2>$. The corresponding transition is shown by red arrow. The gain system is also characterized by the relaxation processes: from level $|2>$ to level $|1>$ with the rate $\gamma_{21}$ and from level $|1>$ to level $|0>$ with the rate $\gamma_{10}$. The frequency of the plasmonic dipole mode of the metal nanosphere is $\omega_{\mathrm{sp}}$.  This mode is coupled to the inter-level transition $|1>\rightarrow |0>$ with the frequency $\omega_{10}\approx \omega _{sp}$. 
		 }
	\label{3leveldiagram}
\end{figure}

\begin{figure}
	\begin{center}
		\includegraphics[width=0.40\textwidth]{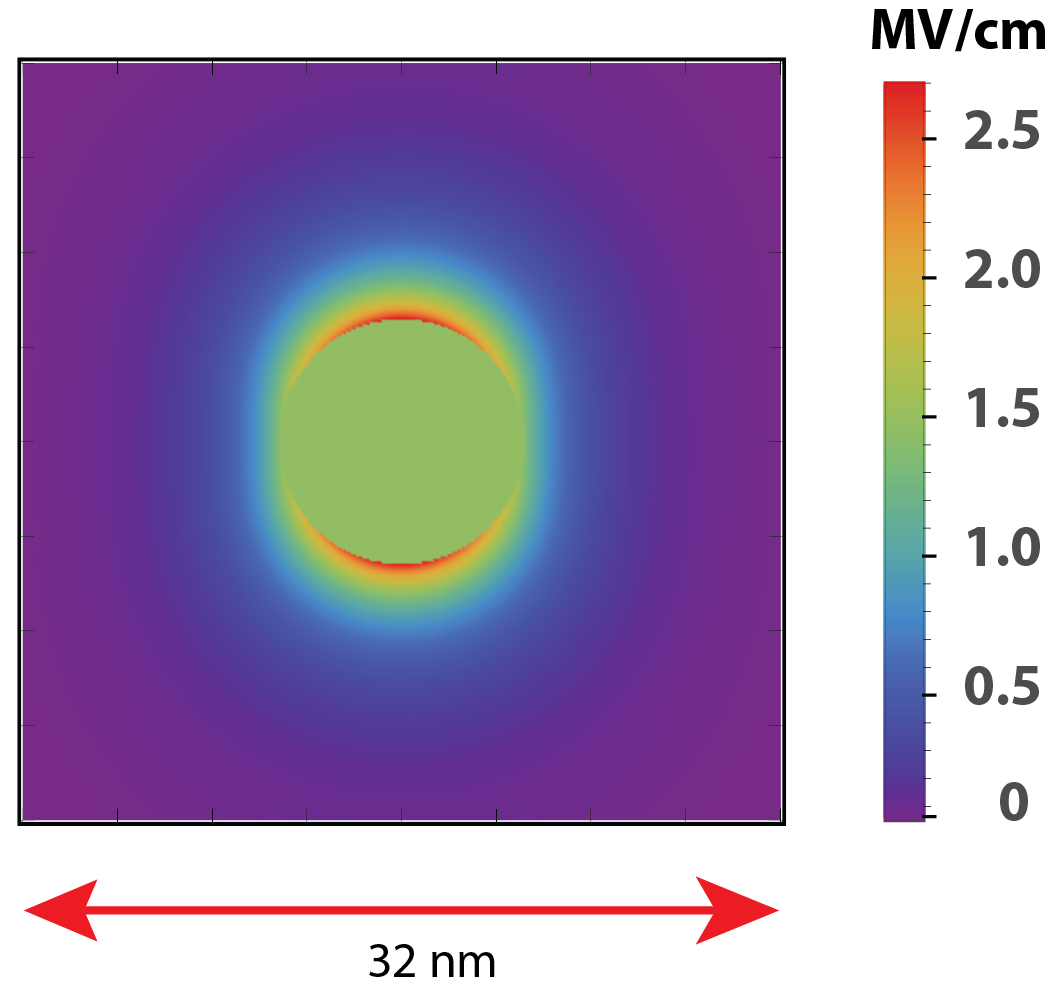}
	\end{center}
	\caption{Electric field of the dipole mode of the spaser. The diameter of the spaser is 32 nm with a 10 nm diameter metal core inside it.  }
	\label{modes}
\end{figure}

\section{Model and Main Equations}

We consider a gain medium consisting of three energy levels with the corresponding populations $n_0$, $n_1$ and $n_2$. This approach is similar to the previously studied 2-level  \cite{stockman10_Nanoscale_quantum_generator} system, but with an additional energy level added to account for the finite relaxation rate of electron population from the highest level to the second excited level.  Many other noticeable works have been done previously to deal with the effects of multi-level\cite{song18_Three_level_spaser,zhong2013all} on the spasing process. However, this paper describes an elegant way to account for coherent processes in gain, which makes our findings in close appropriation with the experimental ones as discussed below. 
The general schematics of transition within the system is shown in  
Fig. \ref{3leveldiagram}.

The gain system is pumped by an external light, which excites the system from the ground state $|0>$ to the second excited state $|2>$ with the transition (gain) rate $g$. The excited states of the system are also characterized by  relaxation rates $\gamma_{21}$ and $\gamma_{10}$, which represent the transitions $|2>\rightarrow |1>$ and $|1>\rightarrow |0>$ respectively. The gain medium is coupled to the plasmonic system through the field-dipole interaction and it is at the almost resonant condition with the frequency $\omega_{12}$ is close to the surface plasmon frequency, $\omega_{\mathrm{sp}}$, see  Fig. \ref{3leveldiagram}.
For more than two mediums present in the system, we use the standard Laplace equation approach as adopted in a popular spaser research\cite{galanzha17_Biological_probe} to calculate the field of the Localized Surface Plasmons. In the spherical system, we consider the electric potential of the dipole mode to be of the form 
\begin{align}
	\phi_{i}(\mathbf{r})=\left( \frac{a_i}{r^2} + b_i r \right) Y_{10}(\mathbf{r}),
\end{align}
where $i$ labels the medium ($i=1,2,3$), $a_i$ and $b_i$ are coefficients corresponding to medium $i$, and $Y_{10}(\mathbf{r})$ is a spherical harmonics  $(l=1$ and $m=0)$.
The Maxwell's continuity equations across the interfaces of the layers are given by the following expressions
\begin{align}
	\phi_{i}(\mathbf{r_i})&=\phi_{i+1}(\mathbf{r_i})\label{boundary1},\\
	\epsilon_i \frac{\partial{}}{\partial{r}}\phi_{i}(\mathbf{r_i})&=\epsilon_{i+1}\frac{\partial{}}{\partial{r}}\phi_{i+1}(\mathbf{r_i})\label{boundary2} ,
\end{align}
where $\epsilon_i $ is the permittivity of medium $i$. For our system, which consists of 3 layers, silver sphere, dye, and water, we solve Eqs. \eqref{boundary1} and \eqref{boundary2} to obtain permittivity of silver ($\epsilon_s$) as a function of permittivities of dye ($\epsilon_d$) and water ($\epsilon_w$), i.e., 
\begin{align}
	\epsilon_s\approx\epsilon_s(\epsilon_d,\epsilon_w).
\end{align}
The frequency of LSPs is then obtained by equating $\epsilon_s$ to the experimental value $\epsilon_{\mathrm{sil}}(\omega)$\cite{johnson72_optical_constant}
\begin{align}
	\epsilon_s=\Re[\epsilon_{\mathrm{sil}}(\omega_{\mathrm{sp}})]
\end{align}
%Then the SP frequency, $\omega_n$, can be found from the following expression
%\begin{align}
%s_n	=\Re[s(\omega)]\biggr\rvert_{\omega=\omega_n} .
%\end{align}
The plasmon dipole mode creates a highly localized dipole-field as shown in the cross-section diagram in Fig.\ \ref{modes}. The corresponding operator of electric field\cite{bergman03_SP_amplification_Stimulated,stockman10_Nanoscale_quantum_generator} can be expressed in terms of creation and annihilation operators, $\hat{a}$ and $\hat{a}^*$, of SP,
%\begin {align}
%\mathbf{E}(\mathbf{r},t)&=-\sum_{n}  A_n (\nabla \phi_n(\mathbf{r}) \hat{a}_n  e^{-i\omega t}+
%\nabla \phi_n^{*} (\mathbf{r}) \hat{ a}_n^{*}  e^{i\omega t}),\label{fieldeq}
%\end{align}
\begin {align}
\mathbf{E}(\mathbf{r},t)&=- A_{\mathrm{sp}} (\nabla \phi(\mathbf{r}) \hat{a}  e^{-i\omega t}+
\nabla \phi^{*} (\mathbf{r}) \hat{ a}^{*}  e^{i\omega t}),\label{fieldeq}
\end{align}
where 
\begin{align}
	 A_{\mathrm{sp}}  =\sqrt{\frac{4\pi 
			\hbar}{s_{1} \derivative{\epsilon_{\mathrm{sil}}(\omega)}{\omega}}}\Big|_{\omega=\omega_{\mathrm{sp}}}
\end{align}
Here, the geometrical parameter $s_{1}$ is given by the following expression
%\begin{align}
%	s_{n,i} = \frac{\displaystyle{ \int_{\mathcal{V}_i}  \theta_i(\mathbf{r}) |\nabla \phi_{n,i}(\mathbf{r})|^2 d^3 \mathbf{r}}}{\displaystyle{ \int_{\mathrm{All~Space}}  |\nabla \phi_{n,i}(\mathbf{r})|^2 d^3 \mathbf{r}}}.
%\end{align} 
\begin{align}
	s_{1} = \frac{\displaystyle{ \int_{\mathcal{V}_{metal}} |\nabla \phi_{i}(\mathbf{r})|^2 d^3 \mathbf{r}}}{\displaystyle{ \int_{\mathrm{All~Space}}  |\nabla \phi_{i}(\mathbf{r})|^2 d^3 \mathbf{r}}}.
\end{align}  

The Hamiltonian of a spaser can be expressed in terms of individual Hamiltonians of surface plasmons and the gain medium, and the dipole type interaction  Hamiltonian of the SP and the gain: 
%\begin{equation}
%	\mathcal{\hat{H}}_{total}= \hbar \sum_{n} \omega_n  \hat{ a}_n^{*} \hat{ a}_n+	\mathcal{\hat{H}}_{gain}+\int_{V} \mathbf{E(r,t)}\mathbf{\hat{d}}~d\mathbf{^3 r} .\label{ham3lvl}
%	\end{equation}
\begin{equation}
	\mathcal{\hat{H}}_{total}= \hbar\omega_{\mathrm{sp}}  \hat{ a}^{*} \hat{ a}+	\mathcal{\hat{H}}_{gain}+\int_{V} \mathbf{E(r,t)}\mathbf{\hat{d}}~d\mathbf{^3 r} .\label{ham3lvl}
\end{equation}
Here $V$ is the total volume of the gain medium and $\mathbf{\hat{d}}$ is the transition dipole moment operator of the gain medium. 

In this article, we adopt a quasi-classical approach \cite{stockman10_Nanoscale_quantum_generator,stockman11_nanoplasmonics_present_past_glimpse} to study the properties of SPs, where the operators $\hat{a}$ and $\hat{ a}^{*}$ are treated as classical variables represented in the form of a time dependent variable  $\hat{a}= a_{0} e^{-i \omega t}$ with $a_{0}$ being the slow varying amplitude. Then, the number of SPs in a given mode then can be written as $N_n=\abs{a_{0}}^2$.

 Below we assume that the corresponding dipole matrix elements are nonzero only between the levels $|0>$ and $|1>$ of the gain system, i.e., only the transitions between these levels can generate SPs. These interactions can be also 
characterized by the Rabi frequency\cite{knight80_Rabi_frequency}, which is given by the following expression 
\begin{align}
	\Omega_{10}(\mathbf{r},t)=\frac{\mathbf{E}(\mathbf{r},t)\mathbf{d_{01}}}{\hbar},
\end{align}
where $\mathbf{d_{01}} =<0| \mathbf{\hat{d}} |1>$.

%For a case of the uniform field the last integral of above equation \ref{ham3lvl} can be written as:
We describe the gain system within the density matrix approach with the corresponding equation of motion 
\begin{equation}
	i \hbar \dot{\mathcal{\hat{\rho}}}(\mathbf{r,t}) =[\mathcal{\hat{\rho}}(\mathbf{r,t}),\mathcal{\hat{H}}],
	\label{density1}
\end{equation}
where $\mathcal{\hat{\rho}}$ is the density matrix of three level gain system. 
Using Rotating Wave Approximation(RWA), we can  express $\mathcal{\hat{\rho}}(\mathbf{r,t})$ as slow-varying diagonal terms and fast varying non-diagonal terms with frequency $\omega \approx \omega_{\mathrm{sp}}$.
\begin{align}
	\mathcal{\hat{\rho}}(\mathbf{r})& =\left(
	\begin{array}{ccc}
		{\rho}_{22}(\mathbf{r},t) & 0 & 0 \\
		0 & 	{\rho}_{11}(\mathbf{r},t) & {\rho}_{10}(\mathbf{r},t) e^{i\omega t} \\
		0 &	{\rho}_{01}(\mathbf{r},t)  e^{-i\omega t} &{\rho}_{00}(\mathbf{r},t)
	\end{array}
	\right).
\end{align}

It is convenient to introduce the following notations: $n_0=\rho_{00}$, $n_1=\rho_{11}$ and $n_2=\rho_{22}$. Then equations for the elements of the density matrix, which can be derived from Eq. (\ref{density1}), take the following form
\begin{eqnarray}
	& & \dot{{\rho}}_{10}(\mathbf{r})=[-i(\omega-\omega_{10})-\Gamma_{10}] {\rho}_{10}(\mathbf{r})+\nonumber \\ 
	& & \hspace{3cm}i n_{10}(\mathbf{r}) {\Omega}_{10}^{*}(\mathbf{r}) a_{0}^{*} , \label{eqcoh4l} \\
	& & \dot{n}_{2}=g n_0-\gamma _{21} n_2 , \label{eqn34l}\\
	& & \dot{n}_{1}=-\gamma _{10} n_1+\gamma _{21} n_2-2 \int_{V}d\mathbf{^3 r} \Im\left(\rho _{10}(\mathbf{r}) a_{0} \Omega _{10}(\mathbf{r})\right), \label{eqn24l}\\
	& & \dot{n}_{0}=-g n_0+\gamma _{10} n_1+2 \int_{V}d\mathbf{^3 r} \Im\left(\rho _{10}(\mathbf{r})  a_{0} \Omega _{10}(\mathbf{r})\right) , \label{eqn14l}
\end{eqnarray}
where $\omega_{10}$ is the transition frequency between $|1>$ and $|0>$ levels of the gain medium and  $g$ is the rate of excitation from the 
$|0>$ level to the $|2>$ level by an external pulse. In the above equations, we also introduced the relaxation rates:  polarization relaxation rate $\Gamma_{10}$ and the  spontaneous relaxation rates $\gamma_{10}$ and $\gamma_{21}$ between the corresponding states as indicated by the indices.

The equation of motion for SPs is obtained from Hamiltonian \eqref{ham3lvl} and is given by the following expression
\begin{align}
	\dot{a}&=-a_{0} \gamma_{\mathrm{sp}}(\omega )+ i (\omega-\omega_{\mathrm{sp}}) a_{0}+\nonumber \\ 
	&\hspace{3cm} i  \int_{V}d\mathbf{^3 r} \left(\rho _{10}^*(\mathbf{r}) \Omega _{10}^*(\mathbf{r})\right),\label{eqplas4l}
\end{align} 
where the plasmon relaxation rate $\gamma_{\mathrm{sp}}(\omega)$ is introduced, $\gamma_{\mathrm{sp}}(\omega)=\frac{\Im{\epsilon_{\mathrm{sil}}(\omega)}}{ \frac{\Re \epsilon_{\mathrm{sil}}(\omega)}{\partial \omega}}$. 
Since the transitions between the levels $|0>$ and $|1>$ are due to coupling to the SPs, the corresponding relaxation rate, $\gamma_{10}$, can be expressed as 
\begin{equation} 
	\gamma_{10}= {|\Omega_{10}|}^2\frac{2(\gamma_{\mathrm{sp}}+\Gamma_{10})}{(\omega_{\mathrm{sp}}-\omega_{10})^2+(\gamma_{\mathrm{sp}}+\Gamma_{10})^2} .
\end{equation} 

Equations (\ref{eqcoh4l})-(\ref{eqplas4l}) determine the dynamics of three level spaser. First, we analyze the stationary solution of these equations, i.e., the continuous wave regime of a spaser. In this regime, the  
time derivatives in the left hand sides of Eqs. (\ref{eqcoh4l})-(\ref{eqplas4l}) are zero. It is convenient to introduce the population 
inversions through the following expressions 
\begin{align}
	{n}_{10}={n}_{1}- {n}_{0}\label{eqinv4l}, \\
	{n}_{21}={n}_{2}- {n}_{1}\label{eqinv4l2}.
\end{align} 
Then taking into account that ${n}_{0}+{n}_{1}+ {n}_{2}=1$, we can express populations of different levels in terms of $n_{10}$ as follows
\begin{align}
	n_0&=\frac{\gamma _{21} n_{10}-\gamma _{21}}{2 \gamma _{21}+g},\label{eqnsolve4la} \\
	n_1&= -\frac{-\gamma _{21}-g n_{10}-\gamma _{21} n_{10}}{2 \gamma _{21}+g},\label{eqnsolve4lb}\\
	n_2&= -\frac{g n_{10}-g}{2 \gamma _{21}+g} .\label{eqnsolve4l}
\end{align}
Substituting Eqs. (\ref{eqnsolve4la})-(\ref{eqnsolve4l}) into the system of equations (\ref{eqcoh4l})-(\ref{eqplas4l}) we obtain the following solution of the stationary equations 
\begin{align}
	\rho _{10}(\mathbf{r})= -\frac{\left(a_{0}^* n_{10} \Omega _{10}^*(\mathbf{r})\right)}{i \Gamma_{10}-\omega _s+\omega _{10}}, \label{eqstatrho4l}
\end{align}
\begin{equation}
	n_{10}= \frac{(\omega_s-\omega_{\mathrm{sp}})(\omega _{10}-\omega _s)+\gamma_{\mathrm{sp}} \Gamma_{10}}{V \rho \Omega _{10}^2} , \label{eqsoln2}
\end{equation}
\begin{align}
	N_n=|a_{0}|^2=&\frac{\Gamma_{10}^2+(\omega_s-\omega_{10})^2 }{2 n_{10} \Gamma_{10}} \times \nonumber\\
	&\frac{\gamma _{21} \left(g - \gamma _{10}-n_{10} \left(\gamma _{10}+g\right)\right)-\gamma _{10} g n_{10}}{ \left(2 \gamma _{21}+g\right) \Omega _{10}^2}.\label{Nn}
\end{align}
In the above equations we assumed that the Rabi frequency is constant within the gain medium and the corresponding integrals in Eqs. (\ref{eqn24l}) and (\ref{eqn14l}) can be replaced by $V$. Here the spasing frequency $\omega_s$ is given by the following expression
\begin{align}
	\omega_s=\frac{\omega_{10}\gamma_{\mathrm{sp}}+\Gamma_{10} \omega_{\mathrm{sp}}}{\Gamma_{10} +\gamma_{\mathrm{sp}}}, \label{omegas1}
\end{align} 
where $\omega_{\mathrm{sp}} < \omega_s < \omega_{10}$, and in the absence of detuning, i.e., if $\omega_{\mathrm{sp}} = \omega_{10}$, it is equal to  $\omega_{\mathrm{sp}}$.

From the above expressions we can identify the effect of relaxation rate 
$\gamma_{21}$ on the main spaser characteristics such as population inversion $n_{10}$, spasing frequency $\omega_s$, threshold, and the number of generated plasmons $N_n$. From Eqs. (\ref{eqsoln2}), (\ref{omegas1}) one can see that both the population inversion and the spasing frequency do not depend on $\gamma_{21}$. 

The spasing threshold $g_{th}$ can be found from Eq.\ (\ref{Nn}), where the threshold is determined from the condition $N_n =0$,
\begin{equation}
	g_{th} = \gamma_{10} \frac{ 1+n_{10}  }{ 1- n_{10} - 
		\frac{\gamma_{10}}{\gamma_{21}} n_{10} }.
	\label{gth}
\end{equation} 
Thus the spaser threshold depends on the relaxation rate, $\gamma_{21}$, but since $\gamma_{10}\ll \gamma_{21}$ this dependence is weak. Taking into account that  $\gamma_{10}/\gamma_{21}\ll 1$, we can find the correction to the threshold $g_{th}^{(0)}$ determined by the two-level spaser model
\begin{equation}
	g_{th} \approx g_{th}^{(0)}
	\left(1+   \frac{\gamma_{10}}{\gamma_{21}} \frac{n_{10} }{1- n_{10} }  \right) ,
	\label{gth2}
\end{equation}
where $g_{th}^{(0)} =  \frac{ 1+n_{10}  }{ 1- n_{10}}\gamma_{10}$.

Finite relaxation rate $\gamma_{21}$ also affects the number of generated plasmons, see Eq.\ (\ref{Nn}).  When $\gamma_{21}$ is large enough, i.e., 
$\gamma_{21}\gg g$, we can consider $\frac{1}{\gamma_{21}}$ as a small parameter and find expansion of Eq.\ (\ref{Nn}) in the powers of $\frac{1}{\gamma_{21}}$ as follows 
\begin{align}
	N_n=\frac{\left(\Gamma_{10}^2+(\omega_s-\omega_{10})^2\right) \left(g-\gamma _{10}-n_{10} \left(\gamma _{10}+g\right)\right)}{4  n_{10}\gamma _{10} \Gamma_{10} \Omega _{10}^2}-\nonumber\\
	\frac{g \left(1- n_{10}\right) \left(g-\gamma _{10}\right) \left(\Gamma_{10}{}^2+(\omega_s-\omega_{10})^2\right)}{8  n_{10} \gamma _{10}^2 \Gamma_{10} \Omega _{10}^2}\left(\frac{1}{\gamma_{21}}\right)- \nonumber\\
	\frac{g^2 \left(1- n_{10}\right) \left(\Gamma_{10}^2+(\omega_s-\omega_{10})^2\right)}{16   n_{10}\gamma _{10}^2 \Gamma_{10} \Omega _{10}^2}\left(\frac{1}{\gamma_{21}}\right)^2 . \label{Nnexpanded}
\end{align}
The first term in this expansion is the result for the two-level spaser 
system\cite{stockman10_Nanoscale_quantum_generator}, for which 
$\gamma_{21}\to\infty$. In this case, the number of generated SPs is proportional to the gain, $g$. The second and the third terms give the correction due to finite relaxation rate, $\gamma_{21}$. These terms introduce quadratic dependence on $g$.

\section{Results and Discussions}

\subsection{System and Parameters}

The spaser system is shown schematically in Fig. \ref{two_setup}(a). A solid silver sphere of radius $5.15$ nm is enclosed with a spherical dye layer making the total radius of the system 16 nm. The dielectric permittivity of the dye is $\epsilon_d=2.2$, water is  $\epsilon_w=1.8$ and that of the metal $\epsilon_m(\omega)$ was obtained from the experimental data \cite{johnson72_optical_constant} of optical constants. The gain medium has an electronic band-gap of $\hbar \omega_{10}=3.13$ eV. We have also used a small detuning ($\delta_2=\hbar(\omega_{\mathrm{sp}}-\omega_{10})= 0.05$ eV) to highlight the robustness of the spaser under small frequency mismatch.
The selection of these specific dimensions of the system ensures the matching of SPs frequency $\omega_{\mathrm{sp}}$ to the transition frequency $\omega_{10}$. Also, Additional parameters that is used are: $\hbar\Gamma_{10}= 0.01 $ eV, $d_{10}=1.5 \times 10^{-17} $ esu and the density of chromophores in a gain medium $\rho=1.8 \times 10^{20}$ cm$^{-3}$.

\subsection{Spasing in Continuous Wave (CW) Regime }
\begin{figure}[H]
	\begin{center}
		\includegraphics[width=0.45\textwidth]{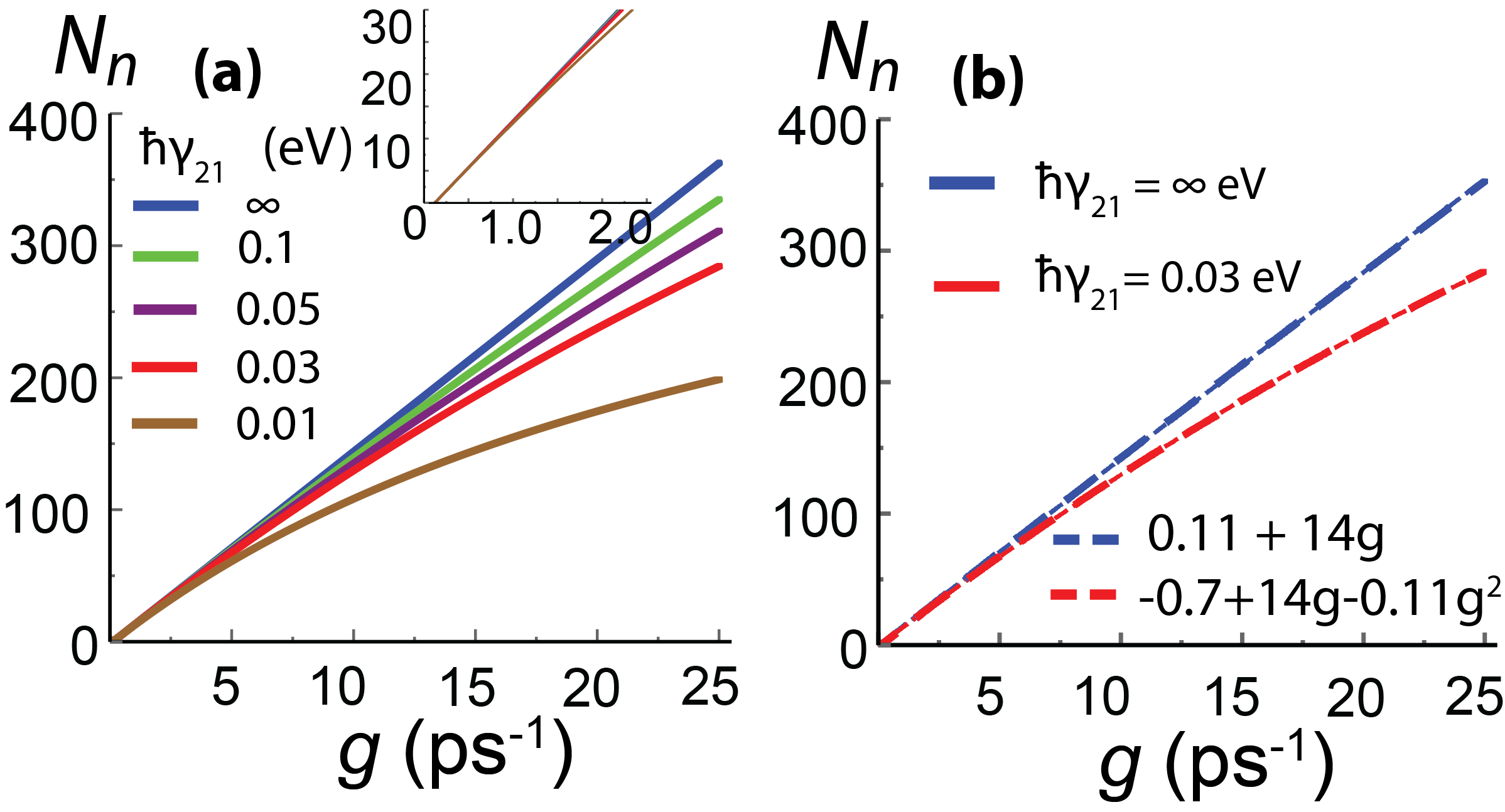}
	\end{center}
	\caption{The number of generated SPs as a function of gain $g$ in the stationary regime. (a) The number of SPs $N_n$ is shown for different values of the relaxation rate $\gamma_{21}$, which characterizes the relaxation from the second excited states of the gain medium to the first excited state. For $\gamma_{21}=\infty $, our model is equivalent to the two-level spaser model. The cropped figure on the top right shows the presence of threshold visible at the lower values of pumping rate (b) The number of SPs as a function of $g$ is shown for two values of $\gamma_{21}$ with the corresponding parabolic fits. While for  $\gamma_{21}=\infty $ the function $N_n(g)$ is a linear function, for  $\hbar \gamma_{21}=0.03 $ eV, it is a parabolic function. }	
	\label{CW}
\end{figure}

In this subsection, we study the stationary solution, which is given by  Eq. (\ref{Nn}), as a function of the pumping rate, $g$: $N_n(g)$. The external optical pulse excites the gain system from the ground level, $|0>$, to the second excited level, $|2>$. If the relaxation from the level 
$|2>$ to first excited level $|1>$ is fast enough, then the spaser system is equivalent to the two-level system\cite{stockman10_Nanoscale_quantum_generator}. In this case, we observe the linear dependence of $N_n$ on the gain rate $g$, see Fig.\ \ref{CW}(a), where the fast relaxation rate, $\gamma_{21}\rightarrow \infty$, is shown by the blue line.

With decreasing the relaxation rate, $\gamma_{21}$, the first excited state, $|1>$, becomes less populated at a given value of the pump rate, $g$, which results in a smaller number of the generated SP. The corresponding dependencies, $N_n(g)$, are shown in Fig.\ \ref{CW}(a) 
for $\gamma_{21}$ in the range of 0.01 eV and 0.1 eV, i.e., for the relaxation time in the range of 6.5 fs and 65 fs. The data clearly show that $N_n$ monotonically decreases with $\gamma_{21}$. For example, at $g=25$ ps$^{-1}$, the number of plasmons decreases by almost a factor of 2 when the relaxation rate decrease from a large value to 0.01 eV. The dependence of $N_n$ on $g$ becomes also parabolic at finite values of $\gamma_{21}$. 

To provide a more clear comparison of the two-level and the three-level spaser systems, we show in Fig.\ \ref{CW}(b) the results for $\gamma_{21}=\infty $ (two-level system) and $\hbar \gamma_{21}=0.03 $ eV, i.e, the corresponding relaxation time is 22 fs, and interpolate them with parabolic dependence. While for the two-level system we have a clear linear dependence on $g$, the three-level system has an extra quadratic term. 
Interestingly, these calculations match the previous experimental findings  \cite{galanzha17_Biological_probe,song18_Three_level_spaser}, where the $N_n$ does not show a linear dependence on the pumping rate, but rather follows a parabolic dependence at a higher pumping rate $g$.

\subsection{Spasing in a Dynamic Regime}
\begin{figure}
	\begin{center}
		\includegraphics[width=0.50\textwidth]{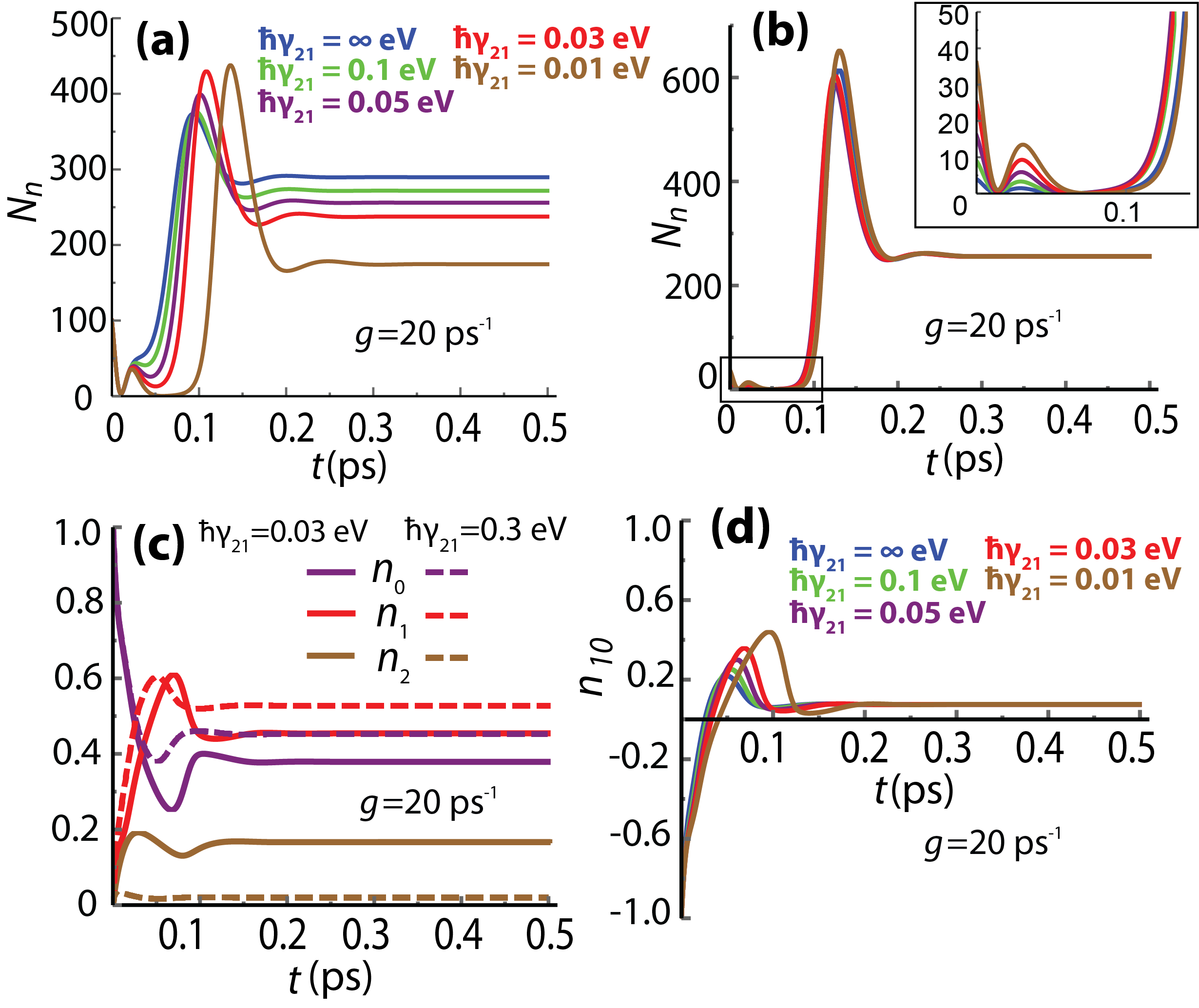}
	\end{center}
	\caption{Illustration of the temporal dynamics of a spaser. 
		(a) The number of generated SPs is shown as a function of time for different values of the relaxation rate $\gamma_{21}$. The gain is 20 ps$^{-1}$. The initial number of plasmons is 100, $N_n(t=0) = 100$. 
		(b) The number of generated SPs is shown as a function of time for different initial numbers of SPs. The gain is 20 ps$^{-1}$ and $\hbar \gamma_{21}=0.05$ eV. (c) Populations $n_2$, $n_1$, $n_0$ of the corresponding levels of the gain medium, $|2>$, $|1>$, $|0>$, are shown as a function of time for two values of the $\gamma_{21}$. (d) The population inversion, $n_{10} = n_1-n_0$, is shown as a function of time. The values of $\gamma_{21}$ are the same as in panel (a). The gain is 20 ps$^{-1}$.  
	}
	\label{temporal}
\end{figure}

The time dynamics of three-level spaser is shown in Fig. \ref{temporal}(a) for different values of $\gamma_{21}$. The initial number of SP, $N_n(t=0)$, does not affect the final value of $N_n$. Thus, we arbitrarily set the initial value of $N_n$  equals 100. The temporal profile of $N_n$ is similar for different values of $\gamma_{21}$ with one difference that with decreasing $\gamma_{21}$ more pronounced oscillation at $t\approx 0.6 $ ps are developed. The final stationary values of $N_n$ follows the results shown in Fig.\ \ref{CW}. 

To illustrate that the stationary value of $N_n$ does not depend on the initial condition, we show in Fig.\ \ref{temporal}(b) the profile $N_n(t)$ for different initial values. The relaxation rate is $\hbar \gamma_{21} = 0.05$ eV. 
The results show that the initial value of $N_n$ only affects the amplitude of oscillations while the stationary solution is independent of $N_n(0)$.

%Spasing is a dynamic process and the final number of SPs is merely a saturated value of the time varying process. The time-evolution of $N_n$ is illustrated in \ref{temporal}.a  for five different parameters of $\gamma_{21}$ as used in the calculations of \ref{CW}. The initial number of SPs in the system doesn't affect the final population of $N_n$, long as it's above the threshold which can be seen in \ref{temporal}.b. Thus, we arbitrarily set the initial value of $N_n=100$  for our other calculations. To ensure the pump rate covers the whole spectrum of effects of the relaxation, we have set $g$= 20 ps$^-1$ as our pump rate. \ref{temporal}.a shows time dependent solutions of $N_n$ for different $\gamma_{21}$ which  converge to numbers from the stationary solution. 

Other characteristics of the spaser dynamics are populations of three levels of the gain medium, $n_0$, $n_1$, and $n_2$. They are shown in Fig. 
\ref{temporal}(c) for fast and slow relaxation rates, where the solid lines correspond to $\hbar\gamma_{21}=0.03$ eV while the dashed lines
correspond to $\hbar\gamma_{21}=0.3$ eV. The data show that with a fast relaxation rate ($\hbar\gamma_{21}=0.3$ eV) the population of the high energy level $|2>$ is almost zero, which corresponds to the limit of a two-level spaser system. Also, as expected, with increasing the relaxation rate, the populations of the ground and the first excited states, $n_0$ and $n_1$, increases while the population of the second excited state, $n_2$, decreases. The population inversion, which is the difference between populations $n_1$ and $n_0$, $n_{10}=n_1-n_0$, is the same for both large and small relaxation rates. To illustrate this property we show in Fig.\ \ref{temporal}(d) the population inversion for different values of relaxation rate $\gamma_{21}$. The values of $\gamma_{21}$ are the same as in Fig.\ \ref{temporal}(a). In all cases, the stationary population inversion does not depend on the relaxation rate, $\gamma_{21}$. This is consistent with expression (\ref{eqsoln2}), the right-hand side of which does not depend on $\gamma_{21}$.

%To study this deviation from the  linear behavior, we also tried to study the change in population in all three states over the time as shown in \ref{temporal}.c . Here, Populations $n_0$,  $n_1$ and  $n_2$ are represented by purple, red and brown colors respectively. The solid line represents the curves for $\hbar\gamma_{21}=0.03$ eV and dashed line for $\hbar\gamma_{21}=0.3$ eV to account for the slow and fast relaxations. We observe that the population  $n_0$ and $n_1$ are higher whereas  $n_2$ is almost 0 for $\hbar\gamma_{21}=0.03$ eV. However, for the lower relaxation value of $\hbar\gamma_{21}=0.03$ eV, it's just the contrast, and we see a significant rise in $n_2$ relative to the prior case. This reflects that there is significant accumulation of electrons in the third level at lower relaxation.

% \ref{temporal}.c, however, shows that the difference of population between $n_0$ and $n_1$ is the equal in both the cases of solid and dashed lines, which further implies the inversion $n_{10}$ stays constant even through  $\hbar\gamma_{21}$ varies. This can clearly be seen from \ref{temporal}.d, which was calculated for all the relaxation cases in \ref{temporal}.a. Thus, our overall process isn't influenced by the inversion directly, but the aggregate effect of $g$, $\gamma_{21}$ and $n_{10}$ as given by equation \eqref{Nnexpanded}.

\section{Conclusion}

Spaser is a unique system where the coupling of the plasmonic system and the gain medium results in the coherent generation of the localized plasmons at the nanoscale. The theories of spaser that are based on the two-level system of gain have their limitation in explaining the spasing behavior at the high pumping rates.
In the present paper, we considered, within a semi-classical approach, a three-level gain system to identify the effects of relaxation between the non-spasing levels on the spaser dynamics. 
%,  which can be instrumental to provide a more descriptive analysis of the spasing process. 
Our results show that the number of generated surface plasmons strongly depends on the relaxation rate $\gamma_{21}$. At large values of $\gamma_{21}$, i.e., fast relaxation, the three-level system converges to
the regular two-level spaser system\cite{stockman10_Nanoscale_quantum_generator} with linear dependence of the number of generated plasmons on the pumping rate. However, at smaller values of  $\gamma_{21}$, the dependence of the number of plasmons on the pumping rate becomes parabolic, which is more pronounced at large pump intensity. Such behavior is consistent with experimental results\cite{galanzha17_Biological_probe,song18_Three_level_spaser}.  Our spaser model can be beneficial in improving the efficacy of existing models and comparison with experimental results.

\section{Acknowledgments}
Major funding was provided by Grant No. DE-FG02-01ER15213
from the Chemical Sciences, Biosciences and Geosciences
Division, Office of Basic Energy Sciences, Office of Science,
US Department of Energy.
Numerical simulations were performed using support by Grant No. DE-SC0007043
from the Materials Sciences and Engineering Division of
the Office of the Basic Energy Sciences, Office of Science,
US Department of Energy.

\bibliography{ref}
\bibliographystyle{ieeetr}

\end{document}